\documentclass[pdflatex,sn-mathphys-num]{sn-jnl}

\usepackage{graphicx}%
\usepackage{multirow}%
\usepackage{amsmath,amssymb,amsfonts}%
\usepackage{amsthm}%
\usepackage{mathrsfs}%
\usepackage[title]{appendix}%
\usepackage{xcolor}%
\usepackage{textcomp}%
\usepackage{manyfoot}%
\usepackage{booktabs}%
\usepackage{algorithm}%
\usepackage{algorithmicx}%
\usepackage{algpseudocode}%
\usepackage{listings}%

\theoremstyle{thmstyleone}%
\theoremstyle{thmstyletwo}%

\theoremstyle{thmstylethree}%

\begin{document}

\title[Article Title]{
Ultrasound-Guided Real-Time Spinal Motion Visualization for Spinal Instability Assessment
}


\author*[1,2]{\fnm{Feng} \sur{Li}}\email{feng.li@tum.de}
\equalcont{These authors contributed equally to this work.}

\author[1,2]{\fnm{Yuan} \sur{Bi}} 
\equalcont{These authors contributed equally to this work.}

\author[1,2]{\fnm{Tianyu} \sur{Song}}

\author[3]{\fnm{Zhongliang} \sur{Jiang}}

\author[1,2]{\fnm{Nassir} \sur{Navab}} 

\affil[1]{\orgdiv{Chair for Computer Aided Medical Procedures and Augmented Reality}, \orgname{Technical University of Munich}, \orgaddress{\city{Munich}, \country{Germany}}}

\affil[2]{\orgname{Munich Center for Machine Learning (MCML)}, \orgaddress{\city{Munich}, \country{Germany}}}

\affil[3]{\orgname{University of Hong Kong}, \orgaddress{\city{Hong Kong}, \country{China}}}

\abstract{
\textbf{Purpose:} Spinal instability is a widespread condition that causes pain, fatigue, and restricted mobility, profoundly affecting patients’ quality of life. In clinical practice, the gold standard for diagnosis is dynamic X-ray imaging. However, X-ray provides only 2D motion information, while 3D modalities such as computed tomography (CT) or cone beam computed tomography (CBCT) cannot efficiently capture motion. Therefore, there is a need for a system capable of visualizing real-time 3D spinal motion while minimizing radiation exposure.

\textbf{Methods:} We propose ultrasound as an auxiliary modality for 3D spine visualization. Due to acoustic limitations, ultrasound captures only the superficial spinal surface. Therefore, the partially compounded ultrasound volume is registered to preoperative 3D imaging. In this study, CBCT provides the neutral spine configuration, while robotic ultrasound acquisition is performed at maximal spinal bending. A kinematic model is applied to the CBCT-derived spine model for coarse registration, followed by ICP for fine registration, with kinematic parameters optimized based on the registration results. Real-time ultrasound motion tracking is then used to estimate continuous 3D spinal motion by interpolating between the neutral and maximally bent states.

\textbf{Results:} The pipeline was evaluated on a bendable 3D-printed lumbar spine phantom. The registration error was 
$1.941 \pm 0.199$ mm and the interpolated spinal motion error was $2.01 \pm 0.309$ mm (median).

\textbf{Conclusion:} The proposed robotic ultrasound framework enables radiation-reduced, real-time 3D visualization of spinal motion, offering a promising 3D alternative to conventional dynamic X-ray imaging for assessing spinal instability.
}


\maketitle

\section{Introduction}\label{sec1}

Spinal instability refers to abnormal motion between adjacent vertebral segments, typically resulting from trauma, neoplastic processes, or degenerative changes~\cite{izzo2013biomechanics, lim2016indications}. Assessing spinal motion is essential for determining whether a spinal segment is stable or unstable~\cite{winn2022novel}, with instability most frequently observed in the lumbar and cervical regions. The current gold standard in clinical practice is dynamic X-ray imaging, in which patients are instructed to bend the spine in different configurations while sequential X-ray images are acquired to evaluate vertebral motion~\cite{esmailiejah2018diagnostic}. However, X-ray remains a two-dimensional modality, lacking 3D perception of spinal movement. Meanwhile, conventional 3D imaging modalities such as CT and CBCT are inherently static. Although prior studies have demonstrated the added value of 4D CT for spine assessment~\cite{buzzatti2021evaluating,yip20224d}, this comes at the cost of increased radiation exposure.
There is a clear need for the development of a system capable of visualizing 3D spinal motion while minimizing radiation exposure.

\par
Recent advances in robotic ultrasound have opened new possibilities for addressing the challenges described above~\cite{bi2024machine,jiang2023robotic,jiang2024intelligent}. Ultrasound is a non-ionizing and real-time imaging modality, and when combined with robotic tracking, it enables three-dimensional ultrasound compounding, providing volumetric perception of anatomical structures~\cite{ungi2020automatic}. Yang \emph{et al.}~\cite{yang2021automatic} introduced a robotic ultrasound system capable of automatically scanning the spine using visual input from an external RGB-D camera. To ensure stable and high-quality ultrasound acquisition during spinal scanning, Wang \emph{et al.}~\cite{wang2023compliant} developed a compliant joint–based adjustment mechanism mounted at the robot end-effector to maintain a constant contact force with the scanning surface. Furthermore, Esteban \emph{et al.}~\cite{esteban2018robotic} demonstrated the intraoperative application of robotic ultrasound by replacing X-ray fluoroscopy guidance during facet joint insertion procedures.

\par
Nonetheless, due to the acoustic properties of ultrasound waves, ultrasound imaging can only capture the superficial surface of the spine, lacking the ability to visualize the entire vertebral body. 
In light of this limitation, previous research has focused on integrating ultrasound with other imaging modalities, such as MRI or CT, which can provide complete visualization of the spine.
Previous researchers have been dedicated to the development of ultrasound with other medical imaging modalities, which can visualize the whole spine properly, such as MRI~\cite{behnami2017model} or CT~\cite{nagpal2014ct}. Following a different approach, Gafencu \emph{et al.}~\cite{gafencu2024shape} trained a shape-completion network to estimate the full vertebral morphology from partial 3D surface observations of the spine. This technique was later integrated into a robotic ultrasound system to achieve real-time visualization of complete vertebrae based on the 3D surfaces provided by robotic ultrasound~\cite{gafencu2025shape}.
However, since shape completion is an ill-posed problem, the network attempts to infer missing anatomical information from limited input, effectively generating the most likely completion based on the training data. As a result, achieving a perfectly accurate reconstruction remains extremely challenging. 
Meanwhile, accurate diagnosis of spinal instability requires plausible geometric fidelity of vertebral structures, as large reconstruction error may compromise diagnostic outcomes. 

\par
In light of these challenges, this work integrates a single preoperative CBCT model with robotic ultrasound as an auxiliary modality to establish a framework for real-time 3D visualization of spinal motion. The preoperative CBCT provides accurate vertebral geometry, while the non-ionizing and real-time ultrasound measurements are used to drive vertebral motion through a kinematic model. In this way, the approach avoids repeated CBCT or CT acquisitions during motion, thereby substantially reducing radiation exposure compared to dynamic 3D radiographic imaging, while overcoming the limited anatomical coverage of ultrasound alone. As a result, each vertebra retains its original shape and size, achieving substantially improved geometric fidelity despite residual registration errors. Moreover, the kinematic model enables continuous 3D visualization, providing richer information compared with 2D X-ray or static CT/CBCT. 
Specifically, a 3D spine model is extracted from the CBCT scan and registered to the partial spinal surface captured by the robotic ultrasound system. By performing robotic ultrasound acquisition while the spine is bent at maximum lateral and flexion positions, the system enables motion-correlated alignment between the ultrasound-observed surface and the preoperative spine model. Furthermore, by optimising the kinematic model parameters after registration to interpolate between the neutral and bent spine configurations under the guidance of real-time 2D ultrasound imaging, which captures the in-plane motion of the vertebrae, a real-time three-dimensional visualization of spinal movement is achieved. This approach provides a radiation-reduced, volumetric, and motion-aware alternative to conventional dynamic X-ray imaging for the functional assessment and diagnosis of spinal instability.

\section{Method}\label{sec2}

\begin{figure}[h]
\centering
\includegraphics[width=0.95\textwidth]{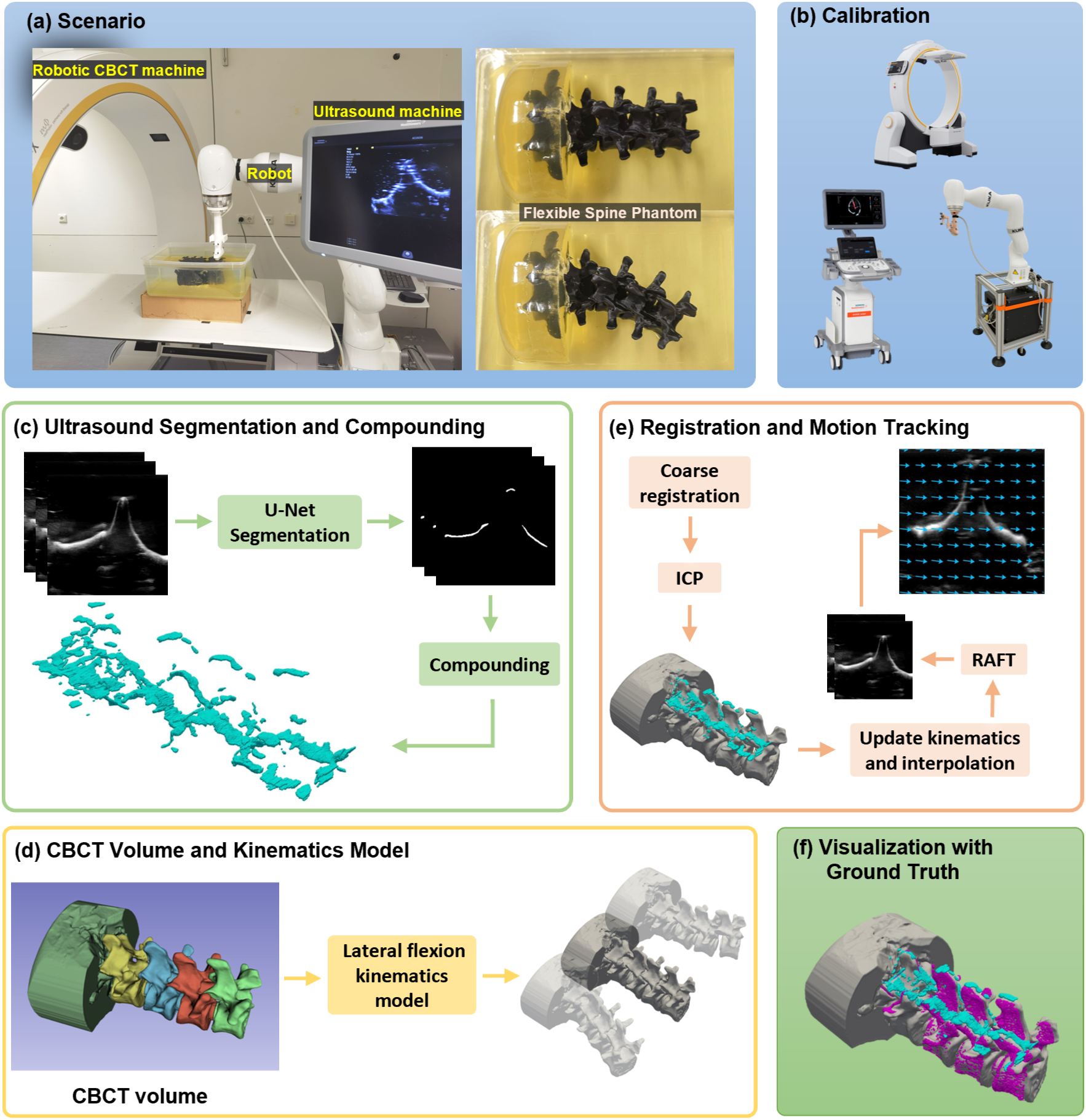}
\caption{Overview of our proposed system.(a) Experimental setup and movable spine phantom. (b) Pre-calibration procedure. (c) Ultrasound acquisition and segmentation at maximal spinal bending with 3D compounding. (d) A kinematic model was initialized and applied to the CBCT volume to simulate spinal motion. (e) Workflow of coarse and fine registration, kinematic parameter optimization, and motion tracking using RAFT optical flow. (f) Comparison of compounded ultrasound, bent CBCT model, and ground truth visualization.}\label{pipeline}
\end{figure}

\subsection{System Overview}\label{subsec1}
\par
Accurate diagnosis of spinal instability requires evaluating spinal motion during bending, with particular emphasis on the spine’s neutral and maximally bent positions. To capture this 3D motion with minimal radiation, our pipeline acquires a single CBCT scan to obtain an accurate 3D spinal model and uses ultrasound, a non-ionizing modality, to capture the spinal surface at maximal bending. The CBCT-derived neutral model and the US-reconstructed bent surface are brought into a shared coordinate frame via cross-system calibration, enabling subsequent registration and motion visualization.

\par
For model-to-surface alignment, we first instantiate a spinal kinematic model to support manual coarse alignment: the operator adjusts L1, while the remaining vertebrae move under kinematic constraints to preserve physiologic behavior; L5 serves as a fixed reference during bending. Starting from this initialization, we apply a rigid Iterative Closest Point (ICP) refinement to register the CBCT model to the compounded ultrasound surface. Moreover, the kinematic parameters were optimized based on the registration results. 

\par
During dynamic acquisition, the ultrasound probe is held on L1 in the axial plane (the segment with the largest displacement). Real-time 2D ultrasound frames provide in-plane vertebral motion, estimated using RAFT optical flow~\cite{teed2020raft}. Given the registered neutral and bent 3D models, we interpolate intermediate poses based on the optimized kinematic model and map the tracked in-plane motion to the corresponding 3D vertebral segment, updating the model in real time to yield a continuous 3D visualization of spinal dynamics. An interactive user interface supports end-to-end visualization of this workflow (Fig.~\ref{pipeline}f). 

Clinically, abnormalities such as spondylolisthesis are primarily assessed by comparing spinal configurations at the neutral and extreme bending positions, where pathological joint behavior is most pronounced. Although the continuous spinal motion visualized in this work is obtained through interpolation and may smooth localized, non-linear dynamics, it remains effective for highlighting atypical vertebral configurations and supporting the identification of spinal instability.
The detailed methodology is described step-by-step in the following sections.

\subsection{Hardware Setup}
\par
The overall system architecture is illustrated in Fig.~\ref{pipeline}(a), where a robotic CBCT system and a robotic ultrasound (US) system operate in a unified and calibrated setup. 
The robotic CBCT system is a commercial device (LoopX, medPhoton, Austria) equipped with an integrated optical tracking camera mounted on the gantry for real-time position monitoring. The robotic ultrasound system consists of a KUKA LBR iiwa 14 R820 robotic arm (KUKA GmbH, Germany) and a Siemens ACUSON Juniper ultrasound machine (Siemens Healthineers, Germany). A linear ultrasound probe (12L3, Siemens Healthineers, Germany) is rigidly mounted to the robot’s end-effector using a custom 3D-printed probe holder. 

\par
To enable convenient data collection while avoiding additional radiation risk, a spine phantom was designed for experimental validation. The phantom consists of: 1) a 3D-printed flexible spine model, a low-cost and reproducible phantom design that has been shown to be effective and easily shareable across clinical institutions, as introduced by West et al.~\cite{west2014development}, and 2) a gelatin medium to simulate soft tissue, following the approach of Jiang et al.~\cite{jiang2021autonomous}. To facilitate smooth spinal motion, both the phantom and gelatin are submerged in water.
To simulate physiological spinal motion, the lumbar vertebrae (L1–L5) are 3D printed individually and assembled using spherical joints, allowing flexible motion at each intervertebral joint. To constrain the motion realistically, the L5 vertebra was fixed using two gelatin blocks, while the upper vertebrae (L1–L4) were left flexible to reproduce continuous bending motion, as illustrated in Fig.~\ref{pipeline}(a). 

\subsection{Pre-Calibration and Data Acquisition}\label{subsec2}

\begin{figure}[h]
\centering
\includegraphics[width=0.9\textwidth]{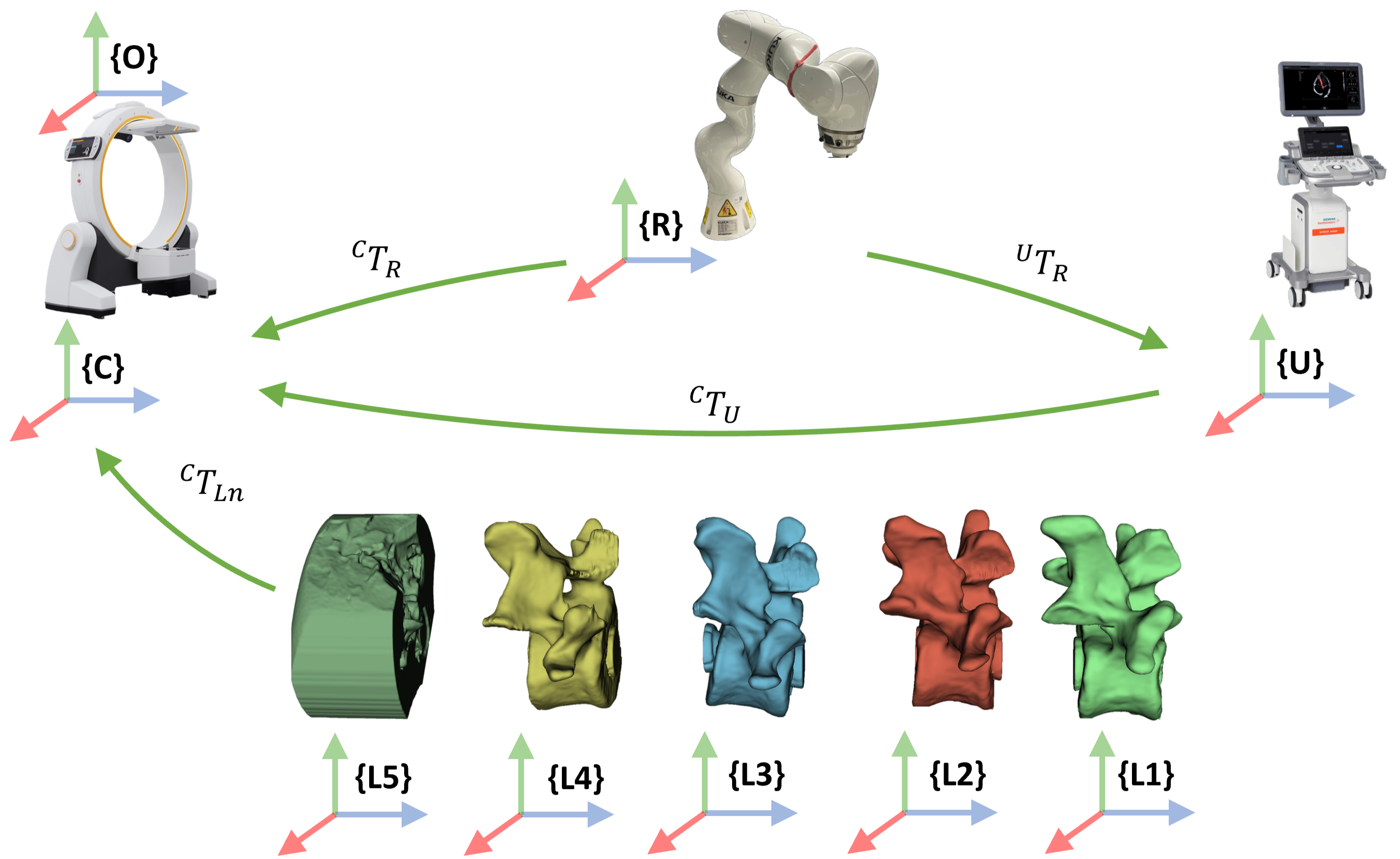}
\caption{Transformation chain of system calibration. This chain illustrates the relationships among the coordinate systems. Specifically, \{R\} denotes the robot frame, \{U\} the ultrasound frame, \{C\} the CBCT volume frame, and \{O\} the LoopX optical tracking camera frame. } \label{transformation}
\end{figure}

In the previous work, a registration framework integrating a robotic CBCT system with a robotic ultrasound (US) platform was proposed, enabling synchronous acquisition of ultrasound images and CBCT slices within the same anatomical region ~\cite{li2025robotic}. The calibration procedure adopted in this study follows the methodology established in that work. As illustrated in Fig.~\ref{transformation}, the goal of the pre-calibration is to establish spatial relationships among all devices in the integrated system. By utilizing the optical tracking system mounted on the LoopX gantry to track the optical marker attached to the robot end-effector, the transformation between the LoopX tracking coordinate system and the robot arm, denoted as $^{O}\mathbf{T}_{R}$, is computed via a hand–eye calibration procedure. The matrix $^{C}\mathbf{T}_{O}$ represents an intrinsic transformation parameter of the robotic CBCT device~\cite{karius2024first}. The ultrasound calibration matrix, $^{U}\mathbf{T}_{R}$, is determined separately following the approach described by Jiang \emph{et al.}~\cite{jiang2021autonomous}. Consequently, the overall transformation between the ultrasound and the CBCT coordinate systems, $^{C}\mathbf{T}_{U}$, can be expressed as:

\begin{equation} 
^{C}\mathbf {T}_{U} = ^{C}\mathbf {T}_{O}(^{R}\mathbf {T}_{O})^{-1}(^{U}\mathbf {T}_{R})^{-1}.
\end{equation}

After completing system calibration, a neutral-spine CBCT volume was first acquired. The lumbar phantom was then manually adjusted to its maximum left lateral, right lateral, and flexion positions
to simulate physiological motion. At each extreme position, the robotic arm was programmed to perform tracked ultrasound scans of the bended spine, providing paired ultrasound volume and CBCT data for subsequent registration. 

\subsection{Ultrasound Segmentation and Compounding}\label{subsec3}

To generate volumetric ultrasound data suitable for registration and motion estimation, the acquired 2D ultrasound frames were first segmented and subsequently compounded into a 3D volume. Specifically, vertebra-wise segmentation was performed to extract the bone surface from each ultrasound image. A U-Net–based segmentation model was adopted from our previous work~\cite{gafencu2024shape}, as it was trained for ultrasound-based spine phantom segmentation, and the phantom used in this study is constructed from the same material, resulting in comparable ultrasound imaging characteristics.
The corresponding robotic pose information for each frame, recorded during scanning, was then utilized to map the segmented 2D slices into a unified 3D spatial coordinate system, enabling consistent volumetric reconstruction. The reconstructed 3D spinal surface is then transformed into the CBCT coordinate system and visualized together with the neutral 3D spinal model in the user interface, allowing the operator to perform manual coarse alignment followed by automatic fine registration.

\subsection{CBCT Volume and Spine Kinematics Model}\label{subsec4}
\par
After CBCT volume acquisition, to obtain a clean and complete representation of the spine, the volume was preprocessed in 3D Slicer using threshold-based segmentation to segment the bone structures. Because the simulated 3D printed spine exhibits strong contrast relative to the surrounding background (gelatin and water), simple thresholding was sufficient to reliably localize the spine. An important step in this process was to separate the spine into individual vertebrae while preserving their spatial alignment relative to the pre-calibrated coordinate system. As illustrated in Fig.~\ref{pipeline} (e) and Fig.~\ref{transformation}, each vertebra was assigned its own local coordinate frame, which serves as the basis for constructing the spinal kinematic model.

\par
In the proposed workflow, spinal lateral and flexion bending were selected as the primary motions to model, though the approach can be readily extended to flexion, extension, and other movement patterns. Each lumbar vertebra (L1–L5) is represented by its own pose transformation matrix, with the neutral spine position defined as the initial reference state. The vertebrae are organized as hierarchically connected nodes from L1 to L5, establishing spatial dependencies along the spinal chain. To simulate the coupled motion of the lumbar spine during bending, a simplified hierarchical kinematic model was implemented: a designated driver vertebra initiates rotation and translation, while its descendant vertebrae inherit progressively attenuated motion according to predefined weighting factors. For example, when L1 is assigned as the driver, L2 follows L1, L3 follows L2, and so forth, with the motion magnitude decreasing sequentially along the chain. This formulation provides a computationally efficient approximation of intersegmental coupling, enabling realistic motion control.

Specifically, we set a value $ s \in [-S_{\max},\, S_{\max}]$ as the input variable that parameterizes bending. Then the corresponding rotation and translation of the driver vertebra can be defined as

\begin{equation}
\theta_1 = \frac{s}{S_{\max}}\, \Theta_{\max} , 
\qquad
d_1 = \frac{ s}{S_{\max}}\,D_{\max},
\end{equation}
where \( \Theta_{\max} \) and \( D_{\max} \) are the maximum rotation and translation, respectively.

If we set the vertebra to rotate with a specific axis $\mathbf{u}$ and translate along the axis $\mathbf{v}$, and $\mathbf{p}\in\mathbb{R}^3$ is the rotation pivot, which is set as the mesh centroid in our implementation. The corresponding rigid-body transformation in homogeneous coordinates is expressed as

\begin{equation}
\mathbf{H}_1(s)
=
\begin{bmatrix}
\mathbf{I} & d_1\,\mathbf{v}\\[3pt]
\mathbf{0}^\top & 1
\end{bmatrix}
\begin{bmatrix}
\mathbf{R}(\mathbf{u},\,\theta_1) & (\mathbf{I}-\mathbf{R}(\mathbf{u},\,\theta_1))\,\mathbf{p}\\[3pt]
\mathbf{0}^\top & 1
\end{bmatrix},
\label{eq:H_driver}
\end{equation}
where $ \mathbf{R}(\mathbf{u},\, \theta_1) $ denotes the $3\times3$ rotation matrix about axis $ \mathbf{u} $ by angle $ \theta_1 $.

For each descendant vertebra at hierarchical depth $ \ell \ge 2 $, the motion increment is proportionally attenuated by a weighting factor $ a_\ell \in [0,1] $:

\begin{equation}
\theta_\ell = a_\ell\,\theta_1, 
\qquad
d_\ell = a_\ell\, d_1.
\end{equation}

The transformation of the \(\ell\)-th vertebra is thus
\begin{equation}
\mathbf{H}_\ell(s)
=
\begin{bmatrix}
\mathbf{I} & d_\ell\,\mathbf{v}\\[3pt]
\mathbf{0}^\top & 1
\end{bmatrix}
\begin{bmatrix}
\mathbf{R}(\mathbf{u},\,\theta_\ell) & (\mathbf{I}-\mathbf{R}(\mathbf{u},\,\theta_\ell))\,\mathbf{p}_\ell\\[3pt]
\mathbf{0}^\top & 1
\end{bmatrix},
\label{eq:H_descendant}
\end{equation}
where $ \mathbf{p}_\ell $ is the pivot of the $ \ell $-th vertebra.

The overall transformation of each vertebra $^{C} \mathbf{T}_\ell$ is calculated as

\begin{equation}
^{C}\mathbf{T}_\ell(s) = \Biggl(\mathbf{H}_\ell(s)\Biggr) \,
^{C} \mathbf{T}_\ell(0),
\label{eq:T_accumulate}
\end{equation}
where $^{C}\mathbf{T}_\ell(0)$ represents the pose of the $ \ell $-th vertebra in its neutral configuration.

\subsection{Registration and Motion Tracking}\label{subsec5}

After kinematic initialization, each vertebra is refined using a constrained, axis-biased ICP algorithm. 
For each vertebra at hierarchical depth $\ell$, we estimate a rigid-body transform 

\begin{equation}
\mathbf{T}_\ell^{*} = \arg\min_{\mathbf{T}\in SE(3)}
\sum_i \|\mathbf{T}\mathbf{x}_i - \mathbf{y}_{j}\|^2,
\label{eq:rigid_transform}
\end{equation}
where $\{\mathbf{x}_i\}$ and $\{\mathbf{y}_j\}$ denote sampled anatomical feature points from the source STL mesh and the target point cloud, respectively. Specifically, the STL mesh is generated from the CBCT volume by applying a kinematic model, while the point cloud is obtained from ultrasound compounding.

Each ICP iteration produces an incremental transform
$\Delta\mathbf{T}_\ell =
\begin{bmatrix}
\mathbf{R(\Theta)} & t\\[3pt]
\mathbf{0}^\top & 1
\end{bmatrix}$
which is constrained by step limits
$ \|\Theta\|\le\Theta_{\max} $, with $\Theta=[\theta_x,\theta_y,\theta_z]^T$ and $ \|\mathbf{t}\|\le t_{\max} $.  
To encourage anatomically consistent motion, translation and rotation are biased toward selected principal axes $ \mathbf{v}$ and $ \mathbf{u}$:

\begin{equation}
\begin{split}
\mathbf{t}' & = 
w_t^{\parallel}(\mathbf{t}\!\cdot\!\mathbf{v})\mathbf{v}
+ w_t^{\perp}\!\left[\mathbf{t}-(\mathbf{t}\!\cdot\!\mathbf{v})\mathbf{v}\right], \\
\mathbf{R}' & =
\mathbf{R}(\mathbf{u},\,w_r^{\parallel}\theta_{\parallel})
\mathbf{R}(\mathbf{u}^{\perp},\,w_r^{\perp}\theta_{\perp})
\end{split}
\label{eq:t_r}
\end{equation}
where $w^{\parallel}$ and $w^{\perp}$ are axis-parallel and orthogonal weights.

The transformation matrix of each vertebra at each update is then given by,

\begin{equation}
\tilde{\mathbf{T}}_\ell^{(k)} = 
\begin{bmatrix}\mathbf{R}'&\mathbf{t}'\\\mathbf{0}^\top&1\end{bmatrix},
\label{eq:stability}
\end{equation}
and accumulated within a bounded trust region,

\begin{equation}
^{C}\mathbf{T}_\ell
\leftarrow
\Big(\prod_{k=1}^{K_\ell}\tilde{\mathbf{T}}_\ell^{(k)}\Big)
^{C}\mathbf{T}_\ell.
\label{eq:trust}
\end{equation}

Following each ICP refinement, the alignment quality is quantified by the Chamfer distance $d_C$ and the symmetric 95th-percentile Hausdorff distance $d_{H95}$.
The attenuation factors $a_\ell$ are automatically updated through an exponentially smoothed rule:
\begin{equation}
a_\ell^{(t+1)} = 
\eta\,a_\ell^{(t)} + (1-\eta)
\big[a_\ell^{(t)} + \alpha\,w_\ell\,\sigma(d_C,d_{H95})\big],
\label{eq:attenuation}
\end{equation}
where $\eta$ is the smoothing constant, $\alpha$ the learning rate, and $w_\ell$ the per-layer tuning weight. $\sigma()$ function encodes whether the registration is “too loose” or “too tight”. $t$ represents the iteration index in optimization. During the parameter-tuning procedure, the attenuation factors $a_\ell$ were updated simultaneously to optimize the model parameters. This adaptive scheme maintains stable registration while gradually refining the kinematic fall-off parameters.

After ICP refinement and the update of attenuation factors, the kinematic model is further optimized to achieve improved motion consistency. All intermediate spine poses between the neutral position and the maximum bending position are interpolated and can be visualized through the developed user interface. During real-time motion acquisition, the ultrasound probe is positioned on L1 in the axial plane, allowing vertebral motion to be captured using optical flow estimation as the spine gradually bends. The RAFT model~\cite{teed2020raft}, a state-of-the-art deep optical flow network pre-trained on large-scale datasets, is employed to compute dense displacement fields between consecutive ultrasound frames. For each frame pair, the optical flow field provides the in-plane displacement, from which the cumulative translation is derived by integrating the displacement across time. This displacement is then normalized relative to the maximum bending distance, yielding a proportional motion parameter. Based on this proportion, the corresponding 3D poses of all vertebrae are obtained through interpolation between the neutral and maximum-bending models.

\section{Results}\label{sec3}
Two experiments were designed to validate the proposed system. The first experiment evaluated the registration accuracy between the 3D spine model extracted from CBCT volume and the compounded ultrasound volume after both coarse alignment and ICP-based fine registration. The second experiment assessed the motion tracking accuracy by first registering the spine at its maximum bending position, then moving it to an intermediate state to quantify the resulting pose estimation error.

\subsection{Registration Accuracy}

After bending the spine phantom, both ultrasound scanning of the bent spine and a CBCT acquisition are performed. Since the CBCT system is spatially calibrated with the robotic ultrasound system, the segmented spine surface from the CBCT scan serves as the ground truth. Following the coarse-to-fine registration process, the registration accuracy is evaluated by comparing the registered mesh (obtained by deforming the spine model extracted from the neutral pose) to the ground-truth point cloud (segmented from the CBCT scan of the bent spine). The nearest-point distance from each vertex in the ground-truth point cloud to the registered mesh was computed, yielding a mean distance of $2.67 \pm 0.313 $ mm and a median distance of $1.94 \pm 0.199$ mm. The achieved registration accuracy falls within the clinically acceptable range for assessing spinal instability, as translational displacements exceeding 4-5 mm are typically considered pathologic in the lumbar spine~\cite{esmailiejah2018diagnostic,panjabi1980basic}.

\begin{figure}[h]
\centering
\includegraphics[width=0.9\textwidth]{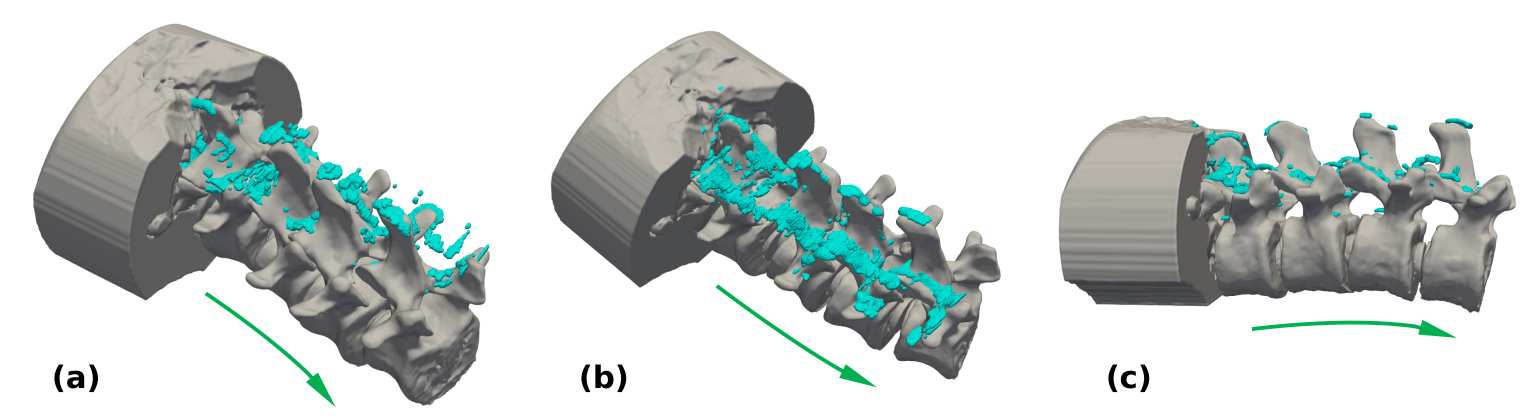}
\caption{Visualization of registration. (a), (b), (c) show the right lateral, left lateral, and flexion bending, respectively. The green line indicates the bending direction in each state. }\label{registration_results}
\end{figure}
Qualitative results of the registration performance are shown in Fig.~\ref{registration_results}. The green point cloud represents the spinal surface reconstructed from the ultrasound volume, while the gray mesh corresponds to the 3D spine model segmented from the neutral CBCT scan. It is clear that the proposed coarse-to-fine registration pipeline can accurately register the 3D spine mesh to the current ultrasound spine surface reconstruction. In some cases, ultrasound compounding is not perfect due to segmentation artifacts. As shown in Fig.~\ref{registration_results}(a), a small number of green points appear slightly displaced, which may affect the accuracy of the ICP step. However, the majority of the points remain well correlated with the spinal surface, causing the ICP optimization to align the plausible vertebral surface rather than being dominated by these outliers.

\begin{table}[ht!]
\centering
\caption{Registration error of individual vertebrae (mm)} \label{tab:acc}
\begin{tabular}{c|cccc}
\toprule
        & L1 & L2 & L3 & L4 \\
\midrule
Mean    &2.04 $\pm$ 0.557 &1.75 $\pm$ 0.126 &1.69 $\pm$ 0.226 &1.59 $\pm$ 0.264 \\
Median  &1.74 $\pm$ 0.554 &1.51 $\pm$ 0.108 &1.42 $\pm$ 0.205 &1.35 $\pm$ 0.250 \\
\bottomrule
\end{tabular}
\end{table}

The registration accuracy of individual vertebrae at maximal bending is summarized in Tab.~\ref{tab:acc}. Accuracy improves slightly from L1 to L4, attributable to reduced motion and increased ultrasound coverage. This means that vertebrae closer to the scan center benefit from a better field of view. In contrast, L4 contains the most usable information due to the linear scanning strategy. 

\subsection{Motion Estimation Accuracy}
To further evaluate the accuracy of motion estimation based on real-time ultrasound, the spine phantom is gradually moved from its neutral position to various bending degrees up to the maximum. The motion of the L1 vertebra is continuously monitored using 2D ultrasound in the axial view. To obtain the ground truth, CBCT scans are acquired at multiple intermediate bending angles. The estimated spine poses, interpolated from the tracked movement of the L1 vertebra observed in 2D ultrasound, are then compared with the corresponding CBCT-derived ground-truth poses to assess motion estimation accuracy. Again, the nearest-point distance from each vertex in the ground-truth point cloud to the registered mesh was computed as the evaluation metric. The results show a mean distance of $3.17 \pm 0.192$ mm and a median distance of $2.24 \pm 0.205$ mm, which are slightly higher than those obtained in the registration experiment.
Moreover, interpolating an intermediate configuration between the neutral and maximally bent positions yields slightly higher accuracy, with a mean distance of $2.64 \pm 0.370$ mm and a median distance of $2.01 \pm 0.309$ mm, indicating improved performance compared with interpolation directly between the neutral and maximal positions.
This degradation in accuracy is expected, as the motion is estimated through interpolation rather than direct measurement. 

\begin{table}[ht!]
\centering
\caption{Motion tracking error of individual vertebrae (mm)} \label{tab:mot}
\begin{tabular}{c|cccc}
\toprule
        & L1 & L2 & L3 & L4 \\
\midrule
Mean    &2.78 $\pm$ 0.131 &2.47 $\pm$ 0.480 &2.27 $\pm$ 0.410 &1.73 $\pm$ 0.283 \\
Median  &2.28 $\pm$ 0.050 &2.19 $\pm$ 0.346 &1.98 $\pm$ 0.279 &1.42 $\pm$ 0.297 \\
\bottomrule
\end{tabular}
\end{table}

The motion tracking accuracy of individual vertebrae is summarized in Tab.~\ref{tab:mot}. Similar to registration accuracy, the accuracy improves slightly from L1 to L4. Overall, the method achieves a spatial accuracy of approximately 3 mm, which falls within the clinically acceptable range of 4–5 mm~\cite{esmailiejah2018diagnostic,panjabi1980basic}.

\section{Discussion and Conclusion}\label{sec12}
This work presents a robotic ultrasound–CBCT fusion framework for radiation-reduced, real-time 3D visualization of spinal motion. By combining a preoperative CBCT-derived spine model with intraoperative ultrasound observations, the system enables volumetric visualization of spinal movement beyond the 2D limits of dynamic X-ray. The proposed coarse-to-fine registration—consisting of a kinematic model, manual alignment, and ICP refinement—achieved accurate registration between modalities on a bendable spine phantom. Real-time motion tracking based on optical flow in ultrasound images further allowed continuous interpolation between neutral and bent poses, providing smooth 3D visualization of vertebral dynamics.

Compared with conventional dynamic X-ray, the method provides 3D anatomical information while minimizing radiation exposure. The robotic setup also improves reproducibility and probe stability, reducing operator dependence. This study has several limitations. First, the quality of ultrasound compounding is influenced by segmentation accuracy. Due to the inherent characteristics of ultrasound imaging, segmentation artifacts may arise and introduce minor spurious structures in the compounded volume. In future studies, particularly when extending the framework to in-vivo data, we plan to incorporate more accurate and advanced spine segmentation models~\cite{ungi2020automatic,bi2023mi} to further improve robustness and visualization quality. Second, due to the limited field of view and the use of 2D ultrasound, the proposed method cannot capture simultaneous dynamic motion across multiple vertebral segments, and out-of-plane rotations may be partially lost. Spinal configurations are acquired at two discrete poses, the neutral and the extreme bending positions, with intermediate motion approximated by interpolation. Nevertheless, abnormal or unstable vertebral behavior can still be identified, as such abnormalities are evident at extreme bending states and remain observable in the interpolated motion. Third, the current evaluation was conducted using a single spine phantom as an initial validation of the proposed framework. Future work will focus on clinical data acquisition and validation in settings closer to clinical practice, while also considering patient comfort and acceptance~\cite{ song2025intelligent}.

In summary, the proposed framework demonstrates the feasibility and potential of robotic ultrasound for functional, real-time 3D assessment of spinal instability.




\backmatter

\bibliography{sn-bibliography}

\end{document}